\documentclass{mem}
\usepackage{txfonts}
\usepackage{balance}
\usepackage{natbib}
\usepackage{graphicx}
\usepackage[a4paper]{hyperref}
\idline{1}{1}

\begin{document}

\title{The high-redshift Universe with the International X-ray Observatory} 

   \subtitle{}

\author{
A. \,Comastri\inst{1}
\and P. \, Ranalli \inst{2,1}
\and R. \, Gilli\inst{1}
\and C. \, Vignali\inst{2,1}
\and M. \, Brusa\inst{3} 
\and F. \, Civano\inst{4}  
         }

  \offprints{A. Comastri}

\institute{
INAF -- Osservatorio Astronomico di Bologna, Via Ranzani 1,
I-40127 Bologna, Italy \email{andrea.comastri@oabo.inaf.it}
\and
Dip. di Astronomia  --
Universit\`a di Bologna, Via Ranzani 1, I-40127 Bologna, Italy
\and
MPE, Giessenbachstrasse 1, D-85748 Garching, Germany  
\and
Center for Astrophysics, 60 Garden Street, 02138 Cambridge MA, USA
}

\authorrunning{Comastri A. }

\titlerunning{High-z SMBH}

\abstract
{We discuss some of the main open issues related to the light--up 
and evolution of the first accreting sources powering  high redshift 
luminous quasars. We discuss the perspectives of future deep 
X--ray surveys with the International X--ray Observatory 
and possible synergies with the Wide Field X-ray Telescope.

}
\maketitle{}

\section{Introduction}

The ``dark ages" of the Universe ended when the UV radiation from the 
first objects reionized the intergalactic medium. This epoch began 
approximately at $6 < z < 14$ and so far has remained essentially 
unexplored.
Investigating the end of the dark ages is extremely important to 
understand structure formation, since during this phase the first
proto--galaxies and the first seed black holes, that would later grow into 
luminous quasars, formed.
Observations of the most distant (z$>6$) quasars (QSOs) thus probe the 
very early growth of supermassive black holes (SMBHs) in the centers of 
massive galaxies.

The high luminosities and broad line widths of the most distant QSOs 
require BH masses above $10^9 M_{\odot}$ \citep{fan01,wil03}.
The formation of such systems within the first 700-800
Myr of cosmic time is a challenge for theoretical models 
\citep[i.e.][]{marta05,li07,beg10}. 
A number of possibilities have been proposed
for the origin of the seed BHs (from Pop III stars to more massive 
objects resulting from the direct collapse of molecular clouds), the 
accretion rate (Eddington limited or super Eddington) and the 
merging rate of dark halos in the early Universe.
Even though suitable combinations of the above parameters may explain 
the presence of massive BHs at $z \sim$ 6, the processes responsible 
for their assembly and light--up  are largely unknown.
The search for and the study of the first QSOs is a key scientific goal 
of future X--ray missions and in particular of the Wide Field X--ray 
Telescope (WFXT; Rosati et al. 2010, this book) and the International 
X--ray Observatory (IXO; White et al. 2010).

\begin{figure*}[t!]
\includegraphics[width=0.45\textwidth, angle=0]{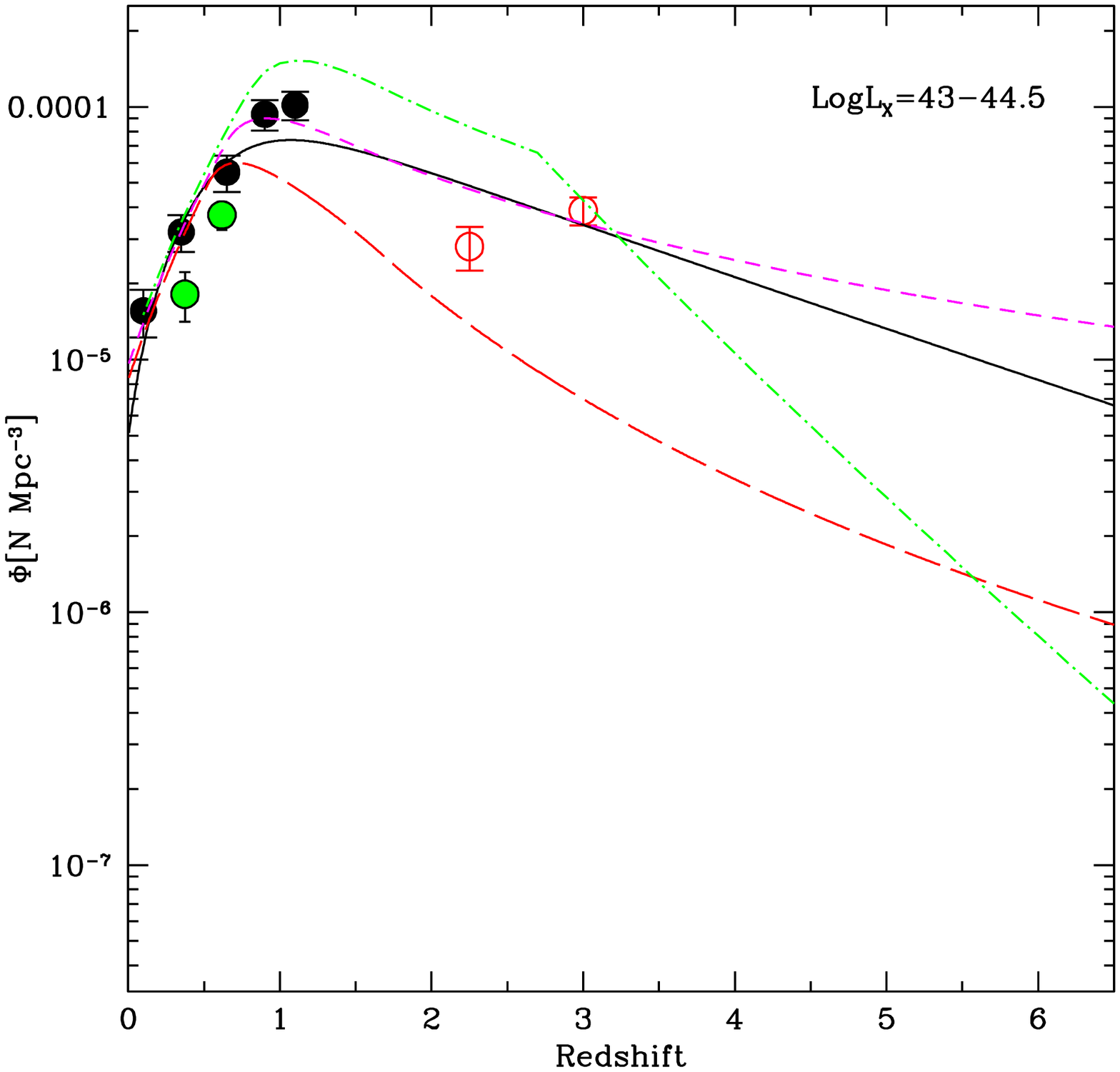}
\hfill
\includegraphics[width=0.45\textwidth, angle=0]{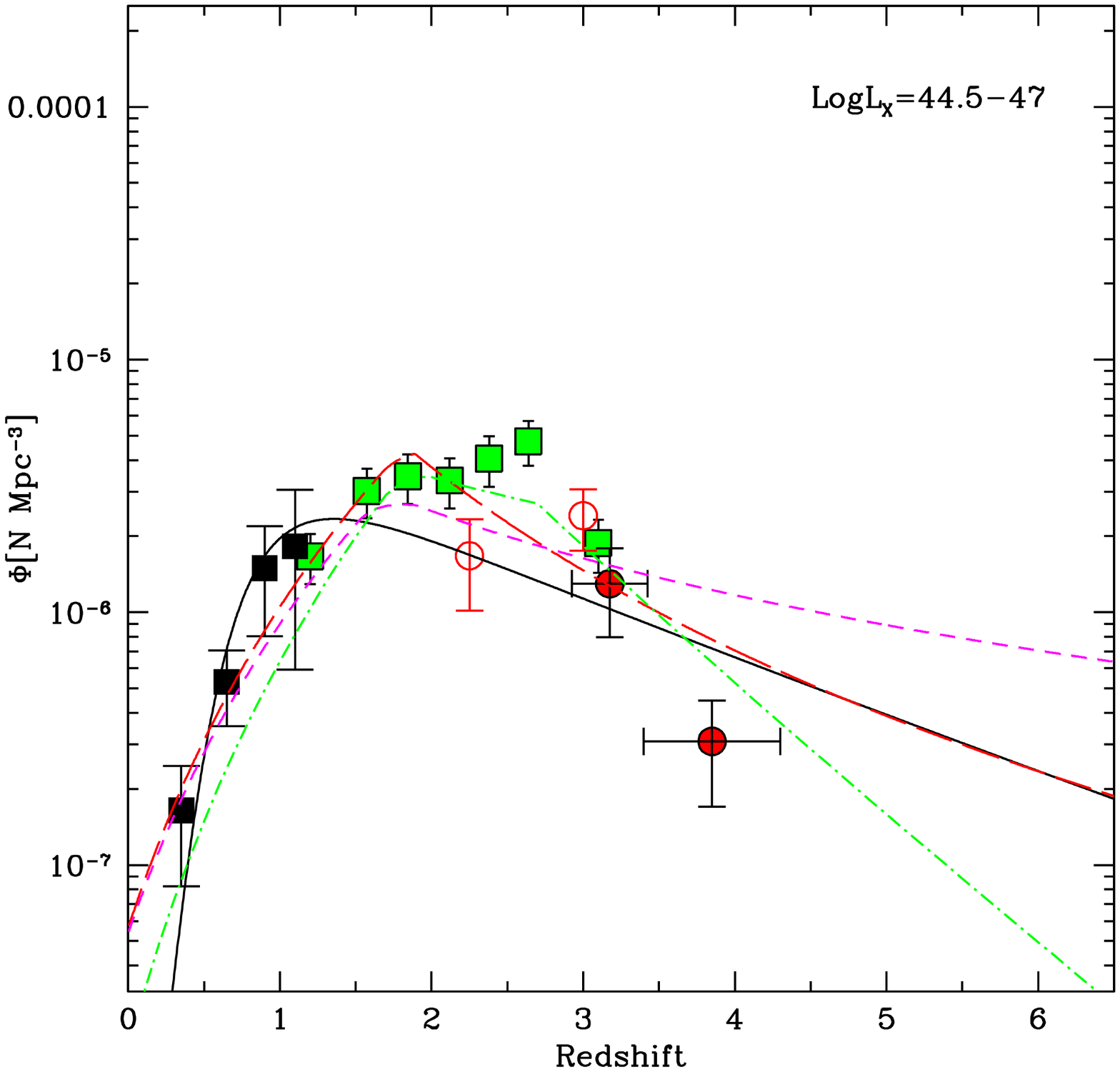}
\caption{\footnotesize 
The AGN number density, as determined by recent X-ray surveys, 
in two 2--10 keV luminosity ranges. {\it Left panel}: log$L_X$=43-44.5;
{\it right panel}: log$L_X$=44.5-47.
Green circles and squares at $z <$ 3 are from the 
XMM-COSMOS survey \citep{bru10}. 
Black circles and squares and open red circles at 2 $< z <$ 3
are from various surveys as described in Aird et al. (2010).
Filled red circles at $z>$ 3 in the right panel are from Brusa et al. (2009).
The best fit results of the various determinations of the X--ray luminosity 
function from Aird et al (2010; black solid line), Ebrero et al. 
(2009; pink dashed curves), Silverman et al. (2008; red dashed curves)
are reported in both panels. 
The green dot--dashed curves represent the expectations of the XRB 
synthesis model of Gilli et al. (2007).
}
\label{eta}
\end{figure*}

\section{Supermassive Black Holes in the early Universe}

Most of the SMBH accretion luminosity is emitted in the optical/UV 
and X--ray bands. As a consequence, optical and X--ray surveys have 
played a major role in the discovery of high--redshift quasars.

\subsection{Optical surveys} 

A major step forward in the study of the high redshift Universe 
was achieved thanks to the Sloan Digital Sky Survey (SDSS).
In a series of papers, Fan and collaborators searched for $z>5.7$ QSOs 
in different releases of the SDSS over an area of about 6600 deg$^2$ 
with a color selection ($i$--band drop out) technique 
(see Fan et al. 2001 for a detailed description).
Their final sample includes 19 objects with $5.7 < z < 6.4$ and 
$M_{1450} < -26$. 
Using the same  technique, the search for $z \simeq$ 6 QSOs was recently 
extended towards lower luminosities ($M_{1450} \sim -24$) 
in the deep SDSS stripe by Jiang et al. (2008,2009), 
allowing the discovery of 12 additional objects.
Thanks to the Canada France High Redshift Quasar Survey (CFHQS),
which covers about 500 deg$^2$ down to fainter magnitudes, 
and applying a similar selection criterion, 19 additional QSOs at $z>$ 5.7
have been found  \citep{wil07,wil09,wil10}.
The number of optically selected QSOs revealed up to $z \sim 6$  
is large enough to determine their luminosity function.
By combining the CFHQS data with the luminous SDSS QSOs, \cite{wil10}
assembled a sample of 40 QSOs to estimate the 5.74 $< z <$ 6.42 
luminosity function over the range 
$-24.7 < M_{1450} < -27.5$. Although the covariance between the slope
and $M_{1450}^{\star}$ luminosity prevents strong constraints to be placed 
on either parameter, there is evidence for the detection of a break
at $M^{\star}_{1450} \simeq -25.1$.  As far as luminous QSOs are concerned, 
an exponential decline ($\propto 10^{-0.43z}$) in their space density from 
$z \sim$ 3 to $z \sim$ 6 has been clearly observed. 
Although SDSS has pushed the redshift record well above $z \sim$ 6
with a statistics good enough to estimate the cosmological space density of high-$z$ QSOs, 
it is crucial to point out that SDSS QSOs are among the brightest 
and most extreme sources in the early Universe 
(in terms of luminosities and BH masses), 
and thus probably not representative of the QSO population at those redshifts.

\subsection{X--ray surveys} 

The {\it Chandra} and XMM sensitivities to faint and hard X--ray sources  
has opened up a new era in the observations of the high redshift Universe. 
The systematic study of high--$z$ ($>$ 4) optically selected 
QSOs was pioneered by 
Brandt and coworkers \citep[i.e.][]{kas00,cris03,she06};
see Brandt et al. 2004 for a review) with pointed snapshot observations, 
and further expanded by cross correlating the SDSS catalogues with the 
XMM archive \citep{you09} and the {\it Champ} project 
\citep{gre09}.
Multifrequency observations 
point towards the absence of evolution in the X--ray spectral index, 
absorption, 
metallicity, emission--line strengths and dust properties
over a broad redshift range 
suggesting that the physical mechanisms powering luminous QSOs
are insensitive to the significant changes on larger scales that occur at
$z \simeq$ 0--6 \citep{maio06,she06}.
Some departures from a self-similar evolution have emerged thanks to 
{\it Spitzer} near infrared observations of a small sample of $z \simeq$ 6 QSOs 
\citep{jia10}. Two out of the 21 $z\simeq$6 QSOs observed by 
{\it Spitzer} do not show any detectable 
emission from hot dust. There is also evidence that hot dust free QSOs 
have the smallest BH masses in the sample (2--3 $\times$ 10$^8 M_{\sun}$)
and are accreting close to their Eddington 
limit, suggesting an early evolutionary stage.  

A major improvement in the study of the cosmological evolution 
of the X--ray emission of moderate to high--$z$ QSOs has been obtained
in the last few years thanks to the large number of deep and medium--deep 
{\it Chandra} and XMM surveys.
Large samples of X--ray selected objects were built, and by now the 
X--ray luminosity function is sampled up to $z \sim$ 3--4, down to 
X--ray luminosities of the order of a few $\times 10^{44}$ erg s$^{-1}$
and to $z \sim$2--3  at lower luminosities (see Fig.~1; 
\citealt{bru09,ebr09,yen09,air10,fon07}).

At higher redshifts, present X--ray observations are highly 
incomplete, being strongly limited by the survey areas. 
The redshift records for X--ray selected QSOs 
are reached in the COSMOS survey ($z=5.41$; Civano et al
2010 ApJ in prep.) and in the CLASXS survey ($z=5.40$; Steffen et al. 2004), 
with only one spectroscopically identified object per survey. 
Deep and ultra--deep {\it Chandra} surveys in the CDFS and CDFN would 
be sensitive to lower X--ray luminosities, but the lack of area 
coverage is much more severe. Also in this case 
only one $z >$ 5 spectroscopically confirmed AGN is found in the CDFN at 
$z$ = 5.19, \citep{bar05}, and none in the CDFS \citep{luo10}.
As a result, the space density of X--ray selected QSOs 
is still unconstrained at $z >$ 5.

\begin{figure*}[t!]
\includegraphics[width=7.5cm]{zgt6_ixo_wb_nolabel_arrows.epsi}
\includegraphics[width=5.5cm,angle=0]{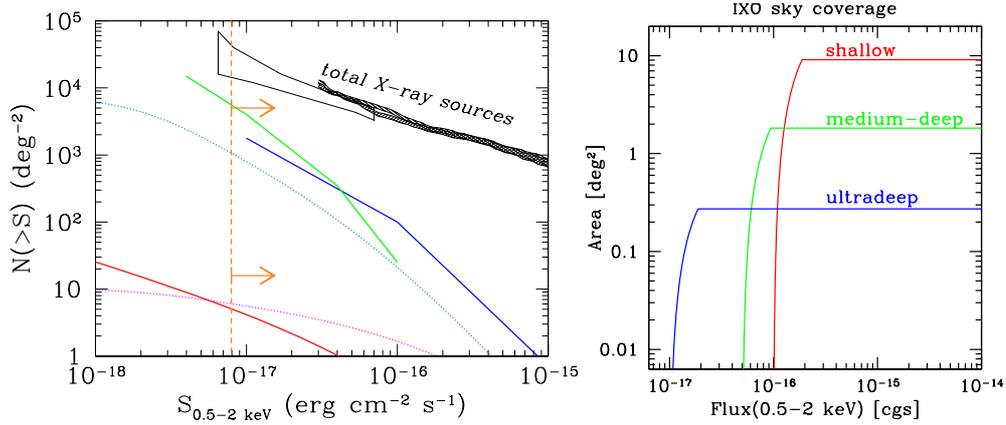}
\caption{\footnotesize {\it Left panel}: 
predicted AGN number counts at $z >$ 6 for various 
semi-analytic models of BH/galaxy formation (solid lines): Salvaterra et al. (2007, green),
Rhook \& Haehnelt (2008, blue) and Marulli et al. (2009, red).
Dotted curves show extrapolations of the luminosity functions adopted 
in the Gilli et al. (2007) model for the synthesis of the XRB (see text for 
details). The vertical orange line represents the IXO confusion limit (reached 
in about 2 Ms).
{\it Right panel}: the sky coverage of a possible ``survey" program 
(see text for details).}
\label{eta}
\end{figure*}

\section{Beyond  $z\sim$ 6} 

The volume and surface density of the first QSOs can be estimated 
following two different approaches. A theoretical one,
by building up semi analytical models (SAMs) which include 
a description as accurate as possible
of the complex physical processes thought to be at work and 
key model parameters such as 
the seed BH mass, the accretion rate and the peak in the 
density field fluctuations above which they can start to form. 
Alternatively, an observational route may be followed.
The best fit luminosity function and evolution, 
as measured by present X--ray surveys or by optical surveys 
can be extrapolated to high redshifts
with some educated guesses on the optical to X--ray luminosity ratio.

Fig.~1 shows the most updated compilation of the X--ray luminosity function (XLF)
parameterized  with a Luminosity Dependent Density Evolution phenomenological model
(LDDE; Silverman et al. 2008; Ebrero et al. 2009) or 
a Luminosity and Density Evolution model (LADE; Aird et al. 2010). 
The predictions based on the AGN synthesis model for the XRB 
of \cite{gil07} are also shown, obtained by including an exponential cut--off
in the XLF evolution at $z$=2.7, which is a good description 
of the observed space density of luminous QSOs in the XMM--COSMOS 
survey \citep{bru09}.

A detailed description of the best fit parameters of each XLF is beyond
the purposes of this paper. Here we would like to stress that 
the extrapolations beyond the redshift range where the XLF was computed 
may differ by more than one order of magnitude. 
Moreover, the faint end of the XLF is poorly determined, even at relatively 
low redshifts (left panel of Fig.~1), making the predictions even 
more uncertain.
The space density of $z>$ 6 QSOs predicted by 
a few recent SAMs \citep{mar09,sal07,rh08} are 
compared with  the extrapolation of the observed XLF 
in the left panel of Fig.~2. 

The lower dotted curve corresponds to the ``decline" model 
(cfr. Fig.~1) which is found to provide a good fit to the 
observed space density of $z \sim$ 3--5 QSOs in the 
{\it Chandra}--COSMOS survey (Civano et al. 2010, in prep.).
The upper dotted curve corresponds to an ad--hoc parameterization
of the XLF high--redshift evolution named ``maxXLF".
The space density of 
low luminosity (log$L_X <$ 44) AGN is kept constant at $z > 4$.
This model maximizes the predicted number of higher redshift 
QSOs, being at the same time in agreement with the observational
results at lower $z$, hence the name ``maxXLF"  
(see Fig.~1 in  Gilli et al. 2010, this volume).   
Different choices for the evolution of the luminosity 
function (cfr. Fig.~1) would correspond to space densities
in between the ``decline" and the ``maxXLF" predictions which 
should then be considered as a conservative and an optimistic 
estimate, respectively. 

\section{Breaking through the first typical AGN}

The bulk of the population of high--z QSOs 
representative of the first accreting objects 
in the Universe is likely to be characterized, 
on average, by relatively low X--ray luminosities and BH masses.
Large area optical and near--infrared surveys such as those foreseen with 
future ground-based facilities and space observatories 
like {\sc panstarrs}, {\sc vista}, {\sc lsst} and 
{\sc jdem/euclid} will discover a large 
number of high-$z$ QSOs. The above surveys will be biased against dust 
obscured QSOs and faint active nuclei for which the host galaxy starlight 
cannot be neglected.

\begin{figure*}[t!]
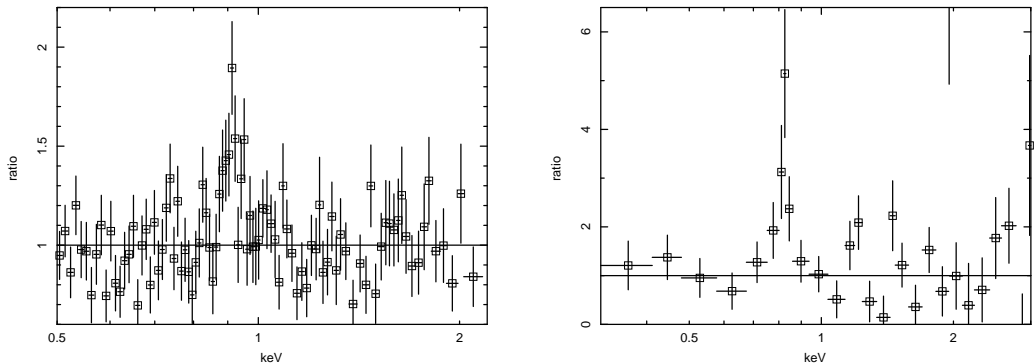

\includegraphics[width=0.35\textwidth, angle=270]{z6_f1e-15_40ev.ps}
\hfill
\includegraphics[width=0.35\textwidth, angle=270]{z7_f1e-16_160eV_obs.ps}
\caption{\footnotesize Residuals wrt a single power law fit to IXO simulations.
{\it Left panel}: a 100 ks simulation, including background, of a z = 6 QSO. 
The line EW of $\sim$ 40 eV observed frame (280 eV rest frame) is clearly seen at $\sim$ 0.9 keV.
{\it Right panel}: a 1 Ms simulation, including background, of a Compton Thick 
AGN at z = 7. The characteristic strong iron line (EW $\sim$ 1.2 keV rest--frame) 
is easily recognizable.
}
\label{eta}
\end{figure*}

X--ray observations can probe  much lower luminosities and high
absorption column densities. 
Moreover, the 
X--ray emission is the telltale of an accreting object. A blind
search for X--ray selected $z\sim$ 5--6 QSOs  would 
require to survey several tens of square degrees to a depth 
comparable to that reached in the {\it Chandra} deep surveys 
(about 0.2 square degrees for a total observing time of 6 Ms) 
and is far beyond the capabilities of current X--ray facilities.

Thanks to its large effective area at 1 keV ($\sim$ 3 m$^2$) and good spatial resolution 
(5 arcsec HPD),
IXO will perform deep surveys reaching limiting fluxes below $10^{-17}$ erg cm$^{-2}$
s$^{-1}$ in the 0.5--2 keV energy range (Fig.~2, left panel). 
The final sensitivity will depend on the actual 
number of faint sources which sets the confusion limit. 
In order to harvest a significant number of moderate luminosity, obscured  
AGN at $z>$ 6 and to cope with the present large uncertainties 
on their predicted space density, a typical multi--cone observing strategy
is mandatory. 

A possible observational strategy for a multi--cone IXO survey with the 
Wide Field Imager (WFI) is shown in the right panel
of Fig.~2. The total observing time invested in cosmological surveys 
would be of the order of 18 Ms broadly split in deep (3 pointings, 2 Ms each) 
medium (20 pointings, 300 ks each) and shallow (100 pointings, 60 ks each) 
corresponding to about 
10 months of observing time assuming an efficiency of
70\%, though it should be noted that 
shallower observations will be performed anyway for a wide 
range of scientific investigations.
\par
The expected number of $z>4$ QSOs will range from a few 
hundreds to more than a thousands, while at $z>$ 6
the predicted number of sources could vary beween a
dozen up to a few hundreds, reflecting  
the uncertainties reported in Fig.~1 and 2.
Taking a median value, the total number of 
high--$z$ QSOs will be sufficient to constrain the
evolution of the faint end of the XLF 
down to $L_X \sim 5 \times$ 
10$^{42}$ erg s$^{-1}$ and up to $z\simeq$6.
IXO will also be able to detect the very first objects 
at $z\sim$ 10, if they exist, and if their 
X--ray luminosity is larger than approximately 
10$^{43}$ erg s$^{-1}$.
IXO deep surveys need to be performed over well  
studied sky areas at longer wavelengths to ease 
the process of source identification and 
redshift determination. The excellent IXO angular 
resolution and sensitivity coupled with the capabilities 
of {\sc jwst}, {\sc alma}, {\sc e-elt} and {\sc tmt} 
will allow to study the assembly of early Black Holes 
and their host galaxies. 

The large IXO collecting area is extremely well suited 
for detailed spectroscopic studies.
According to the currently accepted models 
for the joint evolution of SMBH and their host galaxies 
\citep{hop06,lam10}, the fraction of obscured 
and heavily absorbed or Compton Thick ($N_H > 10^{24}$ cm$^{-2}$) AGN 
is predicted to increase with redshift \citep[i.e.][]{men08}. 
From an observational point of view, the issue of an increased 
fraction of obscured AGN at high redshifts is still debated
(see the discussion in Gilli et al. 2010). 
Irrespective of what the true fraction is, the study of 
heavily obscured AGN at high--$z$ would bear important information 
on the formation and early growth of the first active galaxies.
Primordial, gas and dust rich galaxies, should undergo a phase 
of high obscuration, gas accretion on the central BH and 
vigorous starformation, making deep hard X--ray observations 
an ideal tool for their discovery.
The IXO spectroscopic capabilities are shown in the two panels 
of Fig.~3. The spectrum of a luminous ($L_X \simeq 3\times$ 10$^{44}$ 
erg s$^{-1}$), mildly obscured ($N_H \sim 10^{23}$ cm$^{-2}$) 
QSO at $z$=6 is simulated with a 100 ks WFI exposure.
The 6.4 keV iron line EW is assumed to be 280 eV 
rest--frame. The residuals,
with respect to an absorbed power law fit show a line--like feature 
with an observed frame EW of $\sim$ 40 eV.
In the right panel, the X--ray emission  of a $z=$7, $L_X \sim$ 10$^{43}$ 
erg s$^{-1}$, CT AGN is modeled with a pure reflection spectrum 
plus a strong (EW $\sim$ 1.2 keV rest--frame) iron K$\alpha$ line.  
The residuals of the fit with the same model, without including 
line emission, show that the $K\alpha$ iron line can be used  
to directly measure source redshift with an accuracy 
of $\Delta z \simeq$0.2.  

\section{IXO and WFXT Synergies}

The study of the first SMBH would greatly benefit from joint 
WFXT and IXO surveys. The synergies between the two proposed missions 
are obviously clear.
The WFXT multi-tiered survey strategy is designed to maximize the yield in 
the discovery of high--$z$ AGN. Detailed and quantitative 
estimates are extensively discussed in Gilli et al. (2010, this volume). 
IXO with its superior sensitivity to faint sources would be unique 
for spectroscopic follow--up of the large WFXT samples and to discover 
the faintest AGN at the highest redshifts. 
At the flux of $10^{-15}$ erg cm$^{-2}$ s$^{-1}$, about 
300 AGN at $z>$ 6 are expected in the WFXT medium survey assuming a conservative 
``decline" model. Relatively inexpensive IXO pointings (100 ks each), 
would allow us to obtain spectra of the quality show in the left panel of Fig.~3,
making possible to measure metal abundances and many other QSO physical 
parameters at very high--$z$. 

\bibliographystyle{aa}

\end{document}